\documentclass[reprint,prl,superscriptaddress]{revtex4-1}
\usepackage{bm,color}

\newcommand\beq{\begin{equation}}
\newcommand\eeq{\end{equation}}
\newcommand\beqn{\begin{eqnarray}}
\newcommand\eeqn{\end{eqnarray}}

\newcommand\bl{{\bm{\ell}}}

\newcommand\bL{{\mathbf{L}}}
\newcommand\cl{C_{\ell}}

\newcommand\nn{\nonumber}

\usepackage{graphicx}
\usepackage[large]{subfigure}
\usepackage{amssymb, amsmath}
\usepackage[amssymb]{SIunits}
\usepackage{aas_macros}
\usepackage{natbib}
\renewcommand\section[1]{\emph{#1}.---}

\begin{document}

\title{Detection of the Power Spectrum of Cosmic Microwave Background Lensing by the Atacama Cosmology Telescope}

\author{Sudeep~Das}\affiliation{BCCP, Dept.~of Physics, University of California, Berkeley, CA, USA 94720}\affiliation{Dept.~of Physics,
Princeton University, Princeton, NJ, USA 08544}\affiliation{Dept.~of Astrophysical Sciences, Peyton Hall, 
Princeton University, Princeton, NJ USA 08544}
\author{Blake~D.~Sherwin}\affiliation{Dept.~of Physics,
Princeton University, Princeton, NJ, USA 08544}
\author{Paula~Aguirre}\affiliation{Departamento de Astronom{\'{i}}a y Astrof{\'{i}}sica, Pontific\'{i}a Univ. Cat\'{o}lica,
Casilla 306, Santiago 22, Chile}
\author{John~W.~Appel}\affiliation{Dept.~of Physics,
Princeton University, Princeton, NJ, USA 08544}
\author{J~Richard~Bond}\affiliation{CITA, University of
Toronto, Toronto, ON, Canada M5S 3H8}
\author{C.~Sofia~Carvalho}\affiliation{IPFN, IST, Av.~Rovisco Pais, 1049-001 Lisboa, Portugal}\affiliation{Research Center for Astronomy, Academy of Athens, Soranou Efessiou 4, 11-527 Athens, Greece}
\author{Mark~J.~Devlin}\affiliation{Dept.~of Physics and Astronomy, University of
Pennsylvania, Philadelphia, PA, USA 19104}
\author{Joanna~Dunkley}\affiliation{Dept.~of Astrophysics, Oxford University, Oxford, 
UK OX1 3RH}\affiliation{Dept.~of Physics,
Princeton University, Princeton, NJ, USA 08544}\affiliation{Dept.~of Astrophysical Sciences, Peyton Hall, 
Princeton University, Princeton, NJ USA 08544}
\author{Rolando~D\"{u}nner}\affiliation{Departamento de Astronom{\'{i}}a y Astrof{\'{i}}sica, Pontific\'{i}a Univ. Cat\'{o}lica,
Casilla 306, Santiago 22, Chile}
\author{Thomas~Essinger-Hileman}\affiliation{Dept.~of Physics,
Princeton University, Princeton, NJ, USA 08544}
\author{Joseph~W.~Fowler}\affiliation{NIST Quantum Devices Group, 325
Broadway Mailcode 817.03, Boulder, CO, USA 80305}\affiliation{Dept.~of Physics,
Princeton University, Princeton, NJ, USA 08544}
\author{Amir~Hajian}\affiliation{CITA, University of
Toronto, Toronto, ON, Canada M5S 3H8}\affiliation{Dept.~of Astrophysical Sciences, Peyton Hall, 
Princeton University, Princeton, NJ USA 08544}\affiliation{Dept.~of Physics,
Princeton University, Princeton, NJ, USA 08544}
\author{Mark~Halpern}\affiliation{Dept.~of Physics and Astronomy, University of
British Columbia, Vancouver, BC, Canada V6T 1Z4}
\author{Matthew~Hasselfield}\affiliation{Dept.~of Physics and Astronomy, University of
British Columbia, Vancouver, BC, Canada V6T 1Z4}
\author{Adam~D.~Hincks}\affiliation{Dept.~of Physics,
Princeton University, Princeton, NJ, USA 08544}
\author{Ren\'ee~Hlozek}\affiliation{Dept.~of Astrophysics, Oxford University, Oxford, 
UK OX1 3RH}
\author{Kevin~M.~Huffenberger}\affiliation{Dept.~of Physics, University of Miami, Coral Gables, 
FL, USA 33124}
\author{John~P.~Hughes}\affiliation{Dept.~of Physics and Astronomy, Rutgers, 
The State University of New Jersey, Piscataway, NJ USA 08854-8019}
\author{Kent~D.~Irwin}\affiliation{NIST Quantum Devices Group, 325
Broadway Mailcode 817.03, Boulder, CO, USA 80305}
\author{Jeff~Klein}\affiliation{Dept.~of Physics and Astronomy, University of
Pennsylvania, Philadelphia, PA, USA 19104}
\author{Arthur~Kosowsky}\affiliation{Dept.~of Physics and Astronomy, University of Pittsburgh, 
Pittsburgh, PA, USA 15260}
\author{Robert~H.~Lupton}\affiliation{Dept.~of Astrophysical Sciences, Peyton Hall, 
Princeton University, Princeton, NJ USA 08544}
\author{Tobias~A.~Marriage}\affiliation{Dept.~of Physics and Astronomy, The Johns Hopkins University, Baltimore, MD 21218-2686}\affiliation{Dept.~of Astrophysical Sciences, Peyton Hall, 
Princeton University, Princeton, NJ USA 08544}
\author{Danica~Marsden}\affiliation{Dept.~of Physics and Astronomy, University of
Pennsylvania, Philadelphia, PA, USA 19104}
\author{Felipe~Menanteau}\affiliation{Dept.~of Physics and Astronomy, Rutgers, 
The State University of New Jersey, Piscataway, NJ USA 08854-8019}
\author{Kavilan~Moodley}\affiliation{Astrophysics and Cosmology Research Unit, Univ. of KwaZulu-Natal, Durban, 4041,
South Africa}
\author{Michael~D.~Niemack}\affiliation{NIST Quantum Devices Group, 325
Broadway Mailcode 817.03, Boulder, CO, USA 80305}\affiliation{Dept.~of Physics,
Princeton University, Princeton, NJ, USA 08544}
\author{Michael~R.~Nolta}\affiliation{CITA, University of
Toronto, Toronto, ON, Canada M5S 3H8}
\author{Lyman~A.~Page}\affiliation{Dept.~of Physics,
Princeton University, Princeton, NJ, USA 08544}
\author{Lucas~Parker}\affiliation{Dept.~of Physics,
Princeton University, Princeton, NJ, USA 08544}
\author{Erik~D.~Reese}\affiliation{Dept.~of Physics and Astronomy, University of
Pennsylvania, Philadelphia, PA, USA 19104}
\author{Benjamin~L.~Schmitt}\affiliation{Dept.~of Physics and Astronomy, University of
Pennsylvania, Philadelphia, PA, USA 19104}
\author{Neelima~Sehgal}\affiliation{KIPAC, Stanford
University, Stanford, CA, USA 94305-4085}
\author{Jon~Sievers}\affiliation{CITA, University of
Toronto, Toronto, ON, Canada M5S 3H8}
\author{David~N.~Spergel}\affiliation{Dept.~of Astrophysical Sciences, Peyton Hall, 
Princeton University, Princeton, NJ USA 08544}
\author{Suzanne~T.~Staggs}\affiliation{Dept.~of Physics,
Princeton University, Princeton, NJ, USA 08544}
\author{Daniel~S.~Swetz}\affiliation{Dept.~of Physics and Astronomy, University of
Pennsylvania, Philadelphia, PA, USA 19104}\affiliation{NIST Quantum Devices Group, 325
Broadway Mailcode 817.03, Boulder, CO, USA 80305}
\author{Eric~R.~Switzer}\affiliation{Kavli Institute for Cosmological Physics, 
5620 South Ellis Ave., Chicago, IL, USA 60637}\affiliation{Dept.~of Physics,
Princeton University, Princeton, NJ, USA 08544}
\author{Robert~Thornton}\affiliation{Dept.~of Physics and Astronomy, University of
Pennsylvania, Philadelphia, PA, USA 19104}\affiliation{Dept.~of Physics , West Chester University 
of Pennsylvania, West Chester, PA, USA 19383}
\author{Katerina~Visnjic}\affiliation{Dept.~of Physics,
Princeton University, Princeton, NJ, USA 08544}
\author{Ed~Wollack}\affiliation{Code 553/665, NASA/Goddard Space Flight Center,
Greenbelt, MD, USA 20771}


\begin{abstract}
We report the first detection of the gravitational lensing of the cosmic microwave background through a measurement of the four-point correlation function in the
temperature maps made by the Atacama Cosmology Telescope. We verify our detection by calculating the levels of potential contaminants and performing a number of null tests. The resulting convergence power spectrum at 2-degree angular scales
measures the amplitude of matter density fluctuations on comoving length
scales of around 100 Mpc at redshifts around 0.5  to 3. The measured amplitude of the signal agrees with Lambda Cold Dark Matter cosmology predictions.   Since the amplitude of the convergence power spectrum scales as the square of
the amplitude of the density fluctuations, the 4-sigma detection of the lensing signal measures the amplitude of density fluctuations to 12\%.
\end{abstract}
\maketitle
 
\section{Introduction}
The large-scale distribution of matter deflects the paths of microwave background photons by roughly 3$^\prime$ \cite{cole/efstathiou:1989,*linder:1990,*seljak:1996,*bernadeau:1997}, a scale larger than the $\lesssim 1.4'$ angular resolution of the Atacama Cosmology Telescope (ACT).  This gravitational lensing imprints a distinctive non-Gaussian signature on the temperature pattern of the microwave sky \cite{zaldarriaga/seljak:1999,*okamoto/hu:2003,*lewis/challinor:2006}. Since the cosmic microwave background (CMB) temperature fluctuations are very nearly Gaussian \citep{komatsu/etal:2003, *spergel/etal:2007, *bennett/etal:2011} with a power spectrum now well
characterized by WMAP \citep{larson/etal:2011} and ground-based experiments \cite{brown/etal:2009,*friedman/etal:2009,*reichardt/etal:2009,*sievers/etal:2009,*lueker/etal:2010,das/etal:2011}, measurements of the distinctive four-point correlation function due to lensing yield a direct determination of the integrated mass fluctuations along the line of sight \citep{zaldarriaga/seljak:1999}.  

Previous analyses have detected the lensing signature on the microwave sky through cross-correlations of large-scale structure tracers with WMAP data \citep{smith/etal:2007,hirata/etal:2008}, or seen the signature of lensing in the temperature power spectrum at $\lesssim 3~\sigma$ \citep{reichardt/etal:2009,das/etal:2011}.  Here, we report the first measurement of the lensing signature using only the CMB temperature four-point function and constrain the amplitude of the projected matter power spectrum. 

\section{Data}
ACT is a six-meter telescope operating in the Atacama Desert of Chile at an altitude of 5200 meters. The telescope has three 1024-element arrays of superconducting transition-edge sensing bolometers, one each operating at 148 GHz, 218 GHz, and 277 GHz.  Previous ACT team publications describe the  instrument, observations, and data reduction 
and initial scientific results  \citep{fowler/etal:2010, *swetz/etal:2010, das/etal:2011, dunkley/etal:2010,hajian/etal:2010,*marriage/etal:2010a,
*marriage/etal:2010b, *menanteau/etal:2010, *sehgal/etal:2010b,*hand/etal:2011}. 
The analysis presented here is done on a 324-square-degree stripe of average noise level $\simeq$ 23 $\mu$K-arcmin, made from three seasons of 148~GHz observations of the celestial equator. The region is 
cut into six equally sized (3$\times$18 degree) patches on which we perform  lensing reconstruction separately, and then combine the results with inverse variance weighting.

The ACT temperature maps (made as in \cite{das/etal:2011}) are further processed  to minimize the effects of atmospheric noise and point sources. Temperature modes below $\ell=500$ as well as a `stripe' of width $\ell=180$ along the Fourier axis corresponding to map declination are filtered out to reduce the effects of non-white atmospheric noise and scan-synchronous noise respectively \citep{das/etal:2011}.  Resolved point sources 
with a signal-to-noise (S/N) greater than 5 are identified in a match-filtered map \citep{marriage/etal:2010b}. An ACT beam template scaled to the peak brightness of each of these sources is subtracted from the raw data.  Using an algorithm inspired by the 
CLEAN algorithm \citep{hogbom:1974}, we repeat this filtering, source identification, and subtraction until there are no S/N$>5$ identifications. Because the 148 GHz data also contains temperature decrements from the thermal Sunyaev-Zel'dovich (SZ) effect in galaxy clusters, the entire subtraction algorithm is also run on the negative of the map. The effect of unresolved point sources is minimized by filtering out all data above $\ell=2300$. 

\section{Methods}
Gravitational lensing remaps the CMB temperature fluctuations on the sky: $T(\mathbf{\hat{n}}) = \tilde T(\mathbf{\hat{n}+\bm{\alpha}(\hat{n})})$, where $\bm \alpha(\mathbf{\hat{n}}) $ is the deflection field and unlensed quantities are denoted by a tilde. In this paper, we compute the power spectrum of the convergence field,  $\kappa = \frac{1}{2} \nabla \cdot \bm \alpha$, using
an optimal quadratic estimator \citep{hu/okamoto:2002}:
\beqn
\label{eq:estimator}
 \nn(2 \pi)^2 \delta(\bL&-&\bL') ~\hat C^{\kappa \kappa}_L  =   |N^{\kappa }(\bL)|^2 \int  \frac{d^2\bl}{(2\pi)^2}   \int  \frac{d^2\bl'}{(2\pi)^2}|g(\bl,\bL)|^2  \\
  && \nn\times \left[\vphantom{\int} T^*(\bl)~T^*(\bL-\bl)~ T(\bl')  ~ T(\bL'-\bl')  \right.\\
  && \left. - \left \langle T^*(\bl)~T^*(\bL-\bl)~  T(\bl')  ~T(\bL'-\bl') \right \rangle_\mathrm{Gauss}\vphantom{\int}\right]
\eeqn
where $\bl, \bl', \bL, \bL'$ are coordinates in Fourier space (using the flat-sky approximation), $g$ defines filters that can be tuned to optimize signal-to-noise, $N$ is a normalization, and the second term is the Gaussian part of the four-point function. We will refer to the second term as the ``Gaussian bias'', as it is a Gaussian term one subtracts from the full four-point function to obtain the non-Gaussian lensing signal. 
We normalize the estimator applying the standard formula in   \citep{hu/okamoto:2002,*kesden/cooray/kamionkowski:2003}  using the mean cross-power spectrum estimated from season-splits of the data.   

While the optimal quadratic estimator has the advantage of maximizing the signal-to-noise, an experimental measurement of its amplitude involves subtracting
two large numbers (the full four-point function and the bias). Depending on the quality of data and the relevant length scales, this Gaussian four-point bias term can be up to an order of magnitude larger than the lensing convergence spectrum. As the size of the Gaussian bias term depends sensitively on the CMB temperature power spectrum, foregrounds and  noise, calculating it to sufficient accuracy using the standard simulation or theory approach is very difficult, and can lead to large discrepancies. \citet{smidt/etal:2011} use this standard approach for an analysis of the WMAP data, and report a detection significance larger than expected from Fisher information theory.   
An alternative approach that does not require this subtraction is presented in 
\cite{sherwin/das:prep}.

In this paper, we use the data themselves to obtain a first approximation to the Gaussian bias part of the four-point function, then compute a small correction using Monte Carlo simulations. We first generate multiple randomized versions of the original data map.  The Fourier modes of  these randomized maps have the
 same amplitude as the original map, but with their  phases randomized.  This destroys any non-Gaussian lensing correlation between modes, yet approximately preserves the Gaussian part of the four point function we wish to model. By then averaging the Gaussian biases calculated for many realizations of randomized maps, we obtain a good estimate of the second term in Eq. (1). The small correction we subtract from our estimator (a ``null bias'' at high $\ell$ due to spatially varying noise and window functions) is easily calculated from Monte-Carlo simulations. A similar approach has been suggested by \cite{dvorkin/smith:2009,*hanson/etal:2011}.

\section{Simulations\label{sims}}
We test our lensing reconstruction pipeline by analyzing a large number of simulated lensed and unlensed maps. The simulated maps are obtained by generating Gaussian random fields with the best fit WMAP+ACT temperature power spectrum \citep{das/etal:2011,dunkley/etal:2010}, which includes foreground models, on maps with the ACT pixelization. We then generate lensed maps from these unlensed maps by oversampling the unlensed map to five times finer resolution, and displacing the pixels according to Gaussian random deflection fields realized from an input theory. Finally, we convolve the maps with the ACT beam, and add simulated noise with the same statistical properties as the ACT data, seeded by  season-split difference maps  \citep{das/etal:2011}.

\begin{figure} 
\includegraphics[width=1.\columnwidth]{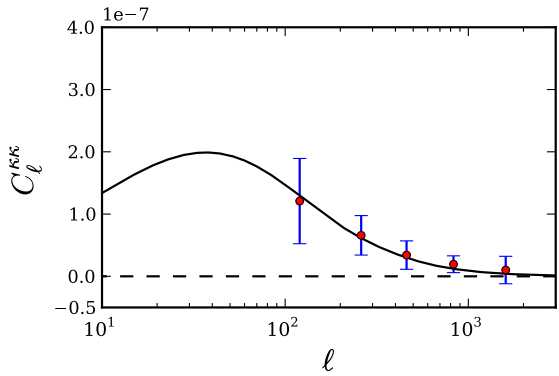}
\caption{
Mean convergence power spectrum (red points) from 480 simulated lensed maps with noise similar to our data. The solid line is the input lensing power spectrum, taken from the best-fit WMAP+ACT cosmological model. Error bars correspond to the scatter of power spectrum values obtained from individual maps.
\label{fig:simulated_cmb}}
\end{figure}

We apply our lensing estimator to 480 simulations of the equatorial ACT temperature map. For each simulated map we estimate the full four-point function and subtract the Gaussian and null bias terms obtained from 15 realizations of the random phase maps. With 15 realizations, the error on the bias contributes $\sim 15\% $ to the total error bars. We thus obtain a mean reconstructed lensing power spectrum, Eq.~(1),  as well as the standard error on each reconstructed point of the power spectrum. The red points in Fig.~\ref{fig:simulated_cmb} show the estimated mean convergence power spectrum from the lensed simulations; it can be seen that the input (theory) convergence power spectrum is reconstructed accurately by our pipeline.

\section{Results}
\begin{figure} 
\includegraphics[width=1.\columnwidth]{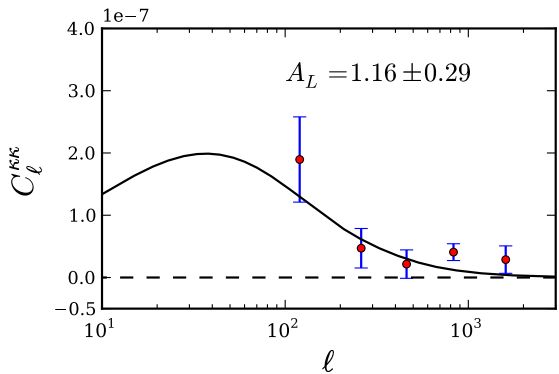}
\caption{Convergence power spectrum (red points) measured from ACT equatorial sky patches. The solid line is the power spectrum from the best-fit WMAP+ACT cosmological model with amplitude $A_L=1$, which is consistent with the measured points. The error bars are from the Monte Carlo simulation results displayed in Fig.~\ref{fig:simulated_cmb}.   The best-fit lensing power spectrum amplitude to our data is $A_L=1.16\pm0.29$ \label{fig:detection}}
\end{figure}
Fig.~\ref{fig:detection} shows the lensing convergence power spectrum estimated from the ACT equatorial data, using the value of the Gaussian term as well as the null bias obtained from the Monte Carlo simulations previously described. The error bars are obtained from the scatter of simulations shown in Fig.~\ref{fig:simulated_cmb}.                                 

Here, we introduce the parameter $A_L$ as a lensing convergence power spectrum amplitude, defined such that $A_L=1$ corresponds to the best-fit WMAP+ACT $\Lambda$CDM model (with $\sigma_8=0.813 $).                                                                                
The reconstructed points are consistent with the theoretical expectation for the convergence power spectrum. From our results we obtain a value of $A_L=1.16 \pm 0.29$, a  4-$\sigma$ detection.  If we restrict our analysis to the first three points, we find $A_L = 0.96 \pm 0.31$. Fitting our five points to the theory, we calculate $\chi^2/\rm{dof}=6.4/4$. Since the lensing kernel has a broad peak at $z \simeq 2$ and a conformal distance of $\simeq 5000$ Mpc, our 4-$\sigma$ detection is a direct  measurement of the amplitude of matter fluctuations at a comoving wavenumber $k \sim 0.02 \mathrm{Mpc}^{-1}$ around this redshift.
\begin{table}
\caption{\label{tab:table1}Reconstructed $\cl^{\kappa\kappa}$ values.}
\resizebox{0.9\columnwidth}{1.3cm}{
\begin{ruledtabular}
\begin{tabular}{lccc}
$\ell$ Range&Central $\ell_b$& $C_b^{\kappa\kappa}$  ($\times 10^{-8}$) & $\sigma(C_b^{\kappa\kappa})$  ($\times 10^{-8}$)\\
\hline
75--150 & 120 &  19.0 & 6.8 \\
150--350  & 260 & 4.7 &  3.2\\
350--550 & 460 & 2.2& 2.3 \\
550--1050 & 830 &4.1 & 1.3  \\
1050--2050 & 1600&2.9& 2.2\\
\end{tabular}
\end{ruledtabular}}
\end{table}

\begin{figure} 
\includegraphics[width=1.\columnwidth]{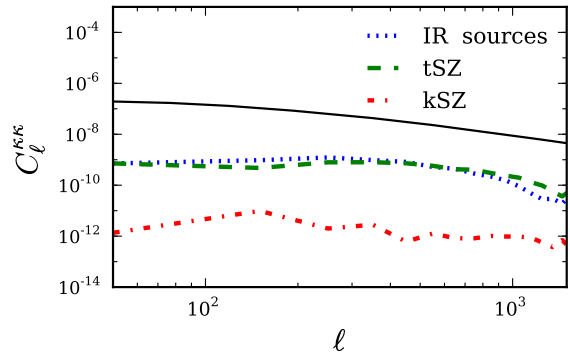}
\caption{Convergence power spectrum for simulated thermal and kinematic SZ maps and point source maps \cite{sehgal/etal:2010a} which are a good fit to the ACT data. Note that we only show the non-Gaussian contribution, as the Gaussian part which is of similar negligible size is automatically included in the subtracted bias generated by phase randomization. The solid line is the convergence power spectrum due to lensing in the best-fit WMAP+ACT cosmological model.
\label{fig:sz}}
\end{figure}

We estimate potential contamination by point sources and SZ clusters by running 
our reconstruction pipeline on simulated patches which contain only IR point sources or only thermal or kinetic SZ signal \citep{sehgal/etal:2010a}, while keeping the filters and the normalization the same as for the data run.  Fig.~\ref{fig:sz}
shows that the estimated spurious convergence power is  at least two orders of magnitude below the predicted signal, due partially to our use of only temperature modes with $\ell<2300$.  We have  also verified that reconstruction on simulated maps containing all foregrounds (unresolved point sources and SZ) and lensed CMB was unbiased. We found no evidence of artifacts in the reconstructed convergence power maps.

\section{Null Tests\label{nulls}}
\begin{figure} 
\includegraphics[width=1.\columnwidth,height=3in]{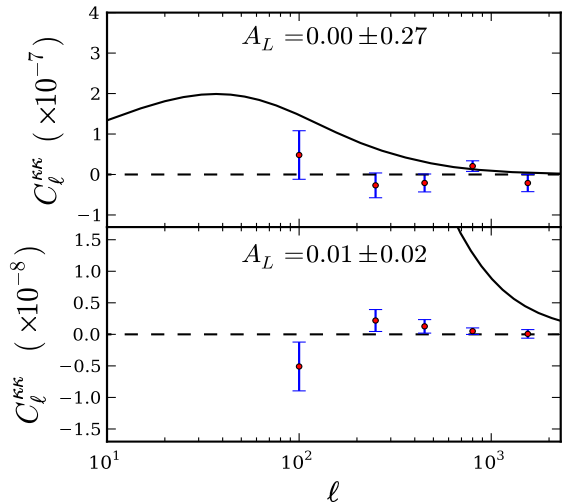}
\caption{\emph{Upper panel:} Mean cross-correlation power spectrum of  convergence fields reconstructed from different sky patches.  The result is consistent with null, as expected. \emph{Lower panel:} Mean convergence power spectrum of noise maps constructed from the difference of half-season patches, which is consistent with a null signal. The error bars in either case are determined from Monte Carlo simulations, and those in the lower panel are much smaller as they do not contain cosmic variance.\label{fig:null_test}}
\end{figure}

We compute a mean cross-correlation power of convergence maps reconstructed from neighboring patches of the data map, which is expected to be zero as these  patches should be uncorrelated. We find  a   $\chi^2/\rm{dof}=5.8/4$ for a fit to zero signal  (Fig.~\ref{fig:null_test}, upper panel).
For the  second null test we construct a noise map for each sky patch by taking the difference of maps made from the first half and second half of the season's data, and run our lensing estimator.   Fig.~\ref{fig:null_test}, lower panel, shows the mean reconstructed convergence power spectrum for these noise-only maps.  Fitting to null we calculate $\chi^2=5.7$ for 4 degrees of freedom. The null test is consistent with zero, showing that the contamination of our lensing reconstruction by noise is  minimal. We also tested our phase randomization scheme by randomizing the phases on a map, using it to reconstruct a convergence map, and cross correlating it with a reconstruction from the same map but with a different phase randomization; our results were consistent with null as expected.

\section{Implications and Conclusions}
We have reported a first detection of the convergence power spectrum of the cosmic microwave background due to gravitational lensing. The inferred amplitude of
the lensing signal is consistent with theoretical expectations of the basic cosmological model.  A detection is also anticipated from the South Pole Telescope team. Data from the Planck satellite \citep{perotto/etal:2010}, and CMB polarization measurements with ACTPol, SPTPol, PolarBear and other next generation experiments \citep{niemack/etal:2010,*2009AIPC.1185..511M,*2008AIPC.1040...66L} will yield even more accurate measurements of CMB lensing. Such measurements are also an important goal for a future polarization satellite mission \citep{smith/etal:2008}. This work  is the first step of an exciting research program.

\begin{acknowledgments}

This work was supported by the U.S. NSF through awards AST-0408698 for the ACT project, and PHY-0355328, AST-0707731 and PIRE-0507768, as well as  by Princeton Univ.~and the Univ.~of Pennsylvania,
RCUK Fellowship(JD),  NASA grant NNX08AH30G (SD, AH and TM), NSERC  PGSD scholarship (ADH),
NSF AST-0546035 and AST-0807790 (AK), NSF Physics Frontier Center grant PHY-0114422 (ES),  KICP Fellowship (ES), SLAC no.\@DE-AC3-76SF00515 (NS), and the BCCP (SD).
Computations were performed on the GPC supercomputer at the SciNet HPC Consortium.  Funding at the PUC from FONDAP, Basal, and the Centre AIUC is acknowledged.
We thank  B.\ Berger, R.\ Escribano, T.\ Evans, D.\ Faber, P.\ Gallardo, A.\ Gomez, M.\ Gordon, D.\ Holtz, M.\ McLaren, W.\ Page, R.\ Plimpton, D.\ Sanchez, O.\ Stryzak, M.\ Uehara, and the Astro-Norte group for assistance with ACT observations.
We thank Thibaut Louis, Oliver Zahn and Duncan Hanson, and  Kendrick Smith for discussions and draft comments.
\end{acknowledgments}
\bibliography{act_lensing_abbrev}

\end{document}